\documentclass[review,number,sort&compress,3p,times]{elsarticle}	

 \usepackage{lineno}
 \usepackage{color}
\usepackage[stable]{footmisc}

\usepackage{amssymb}
\usepackage{subfigure}
\begin{document}
\begin{frontmatter}
\title{Photoemission and optical constant measurements of a Cesium Iodide thin film photocathode}

\author{Triloki, R. Rai, Nikita Gupta, Nabeel F. A. Jammal}
\author {B. K.~Singh\corref{cor}}
\cortext[cor]{Corresponding author}
\ead{bksingh@bhu.ac.in}
\address{High Energy Physics Laboratory, Department of Physics,
  Banaras Hindu University,
  Varanasi 221005 India}

\begin{abstract}

  The performance of cesium iodide as a reflective photocathode is presented. The absolute quantum efficiency of a 500 nm thick film of cesium iodide has been measured in the wavelength range 150 nm to 200 nm. The optical absorbance has been analyzed in the wavelength range 190 nm to 900 nm and the optical band gap energy has been calculated. The dispersion properties were determined from the refractive index using an envelope plot of the transmittance data. The morphological and elemental film composition have been investigated by atomic force microscopy and X-ray photo-electron spectroscopy techniques. 
\end{abstract}

\end{frontmatter}

\section{Introduction}
Various photocathodes are currently used to improve the sensitivity of photon counting or imaging detectors. The choice of photocathode material is determined by the spectral range where the device sensitivity is crucial ~\cite{Tremsin}. Alkali halide photocathodes have been shown to be very efficient photo-converters in the ultraviolet (UV) wavelength range. Cesium Iodide (CsI) is one of the most efficient among them, because CsI photocathode is relatively stable under short exposure to air and has the highest quantum efficiency (QE) among other alkali halide photocathodes ~\cite{breskin}. Therefore it is widely used in many UV-detecting devices ~\cite{breskin}. These devices consist of films of thickness varying from a few nanometers to a few micrometers, depending upon the mode of operation of the photocathode. Therefore it is important to know the absorbance, transmittance and refractive index as a function of wavelength to predict the photoemissive behavior of the photocathode. From the knowledge of these optical constants, the optical band gap of the film can be determined.

In the present work, we have measured the optical absorbance of a 500 nm thick CsI film in the spectral range 190 nm to 900 nm. The transmittance, refractive index and optical band gap energy were estimated from the absorbance data. Photoemission measurements were performed in the wavelength range 150 nm to 200 nm. The surface morphology, studied by atomic force microscopy (AFM) is also reported together with the elemental composition obtained by X-ray photo-electron spectroscopy (XPS).

\section{Experimental Details}

The CsI film was deposited by thermal evaporation (TE) in  a high vacuum ($3\times 10^{-7}$ Torr) evaporation chamber. Prior to deposition,  typical compositions of different residual gases including water vapor inside the evaporation chamber were monitored through a residual gas analyzer (SRS RGA 300 unit). A CsI crystal of high purity (99.999\%) from  Alfa Aesar \footnote{Alfa Aesar: 30 Bond Street, Ward Hill, M A 01835, 800-343-0660}, was placed into a tantalum boat inside the deposition chamber and heated carefully to allow outgassing from the outer surface of the crystals, if any. After appropriate outgassing and melting of CsI crystals, a 500 nm thick CsI film was deposited  on quartz and aluminum (Al) substrates.  The typical deposition rate was 1 nm to 2 nm per second. The thickness of the film as well as the deposition rate were controlled by a quartz crystal thickness monitor (Sycon STM 100). After the film preparation, the vacuum chamber was purged with nitrogen  $(N_{2})$  gas, in order to avoid the interaction of humidity on the prepared CsI film. Immediately after the chamber opening under constant flow of $N_{2}$ gas, the CsI film was placed into a vacuum desiccator and further moved to the characterization setup. 

The photoemission measurement was performed on a 234/302 vacuum ultra violet (VUV) monochromator (see reference [3] for details), in the wavelength range 150 nm to 200 nm. The UV/Vis measurement of CsI films was carried out on a Perkin Elmer UV/Vis spectrometer (Model: $\lambda$ 25)  in the wavelength range 190 nm to 900 nm.  Further, for morphological study, the CsI film deposited on an Al substrate was used for AFM measurement. AFM scanning was done by NEXT ND-MDT atomic force microscope, which provides a high resolution two dimensional (2-D) and three dimensional (3-D) surface image. The elemental composition analysis of the film was carried out by XPS.

\section{Photoemissive properties of CsI}
The absolute quantum efficiency (QE) of the film was measured with the 234/202 VUV monochromator. The absolute QE, which is the ratio of emitted photoelectrons to incident photons, is determined by illuminating the cathode with a photon flux of a given wavelength and the resulting photocurrent measured by a picoammeter (Keithley-6485). The QE measurement of the as-deposited CsI photocathode was  performed in the wavelength range 150 nm to 200 nm, with a step size of 2 nm. From the plot of Figure 1, the maximum QE is about 40\% at 150 nm. The QE was found to decrease with increasing wavelength. The observed QE data are in good agreement with most data reported in the literature~\cite{bksingh}.

\begin{figure}[htb]
  \centering
  \includegraphics*[width=70mm]{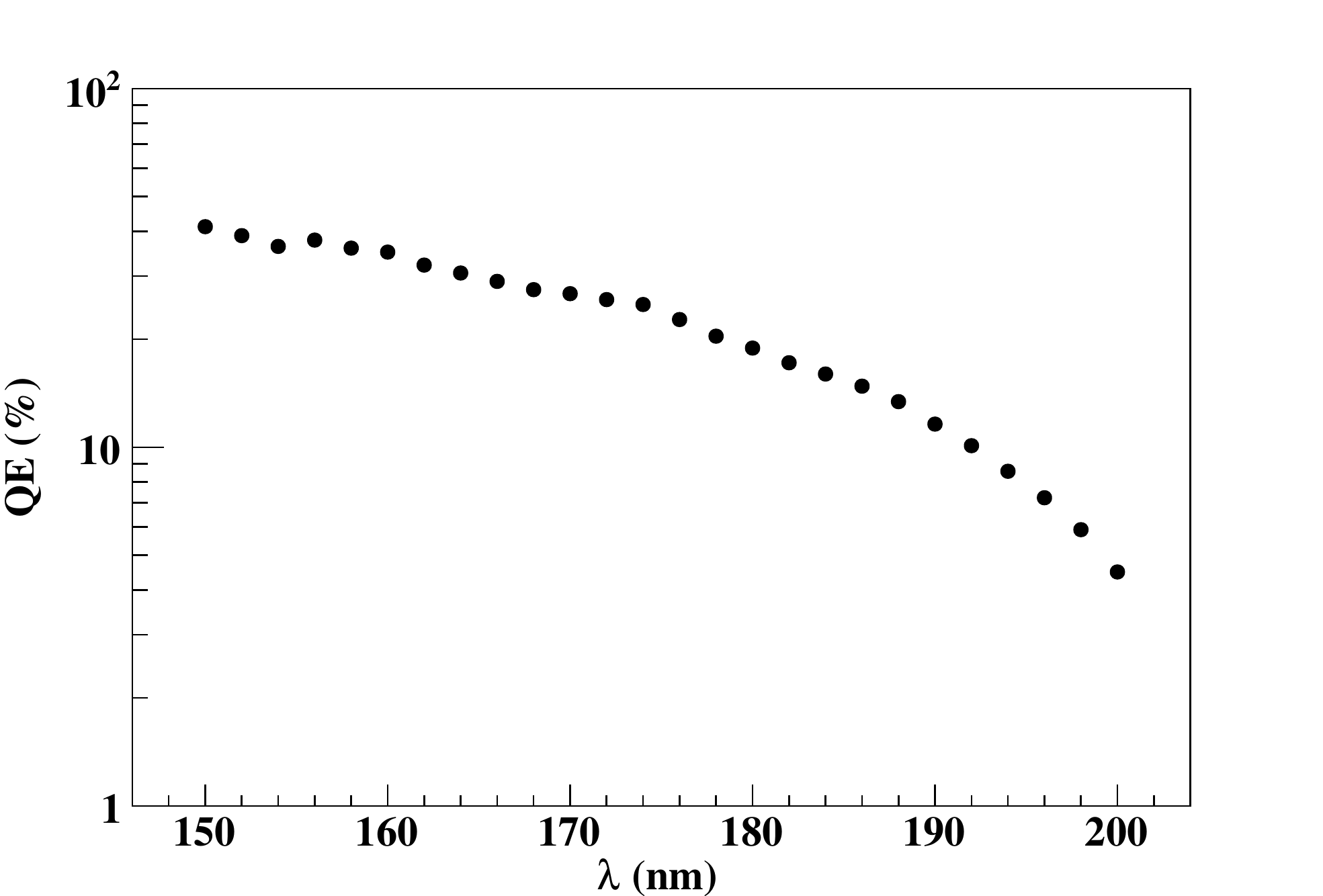}
  \caption{Absolute QE (\%) as a function of  wavelength ($\lambda$) of  a 500 nm thick CsI film photocathode deposited on an Al substrate.}
  \label{l2ea4-f2}
\end{figure}

\section{Optical properties of 500 nm thick CsI film}

\subsection*{Optical absorbance and band gap}

\begin{figure}[hbpt]
  \centering
  \includegraphics*[width=60mm]{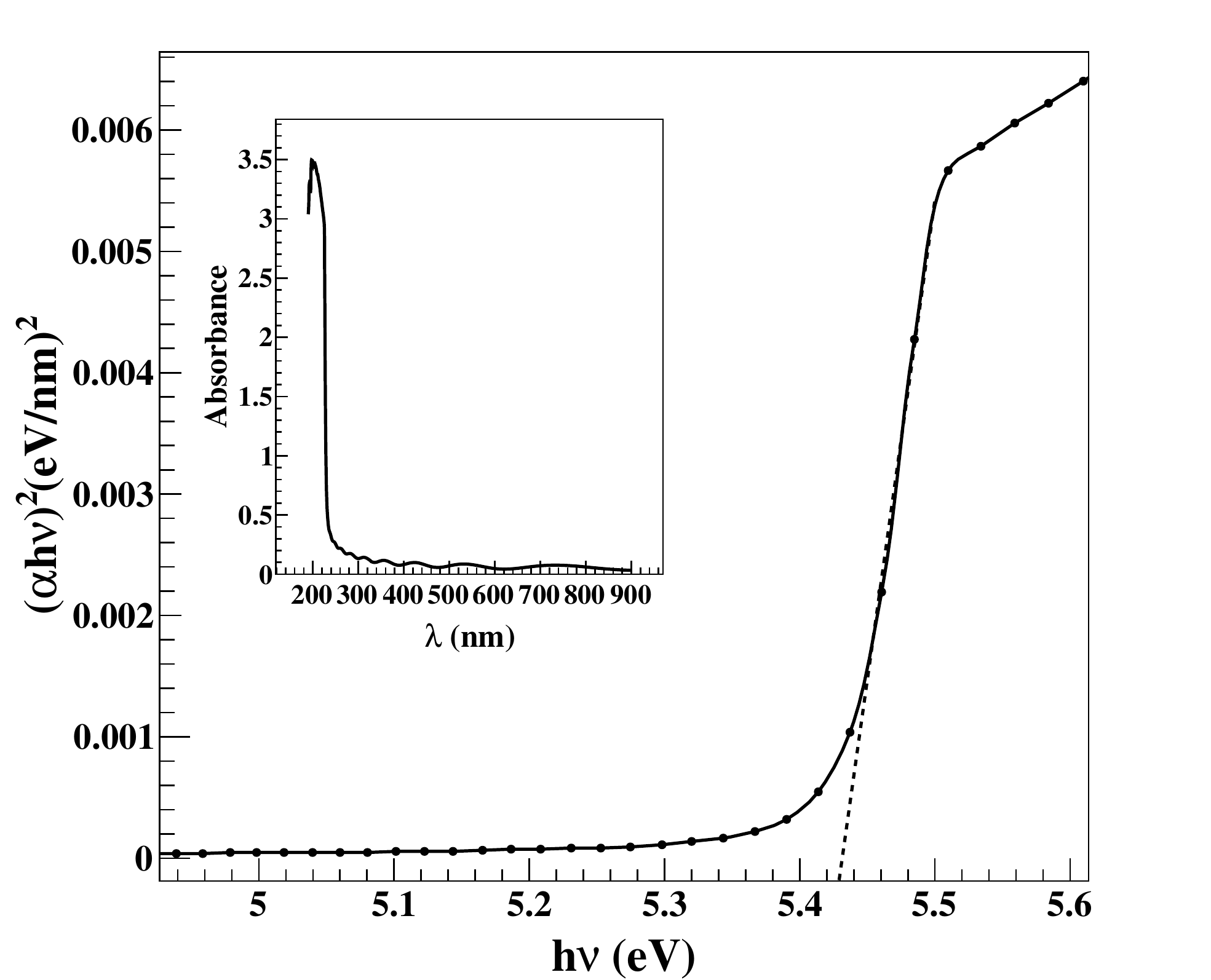}
  \caption{Variation of   $(\alpha h\nu)^{2}$ vs. photon energy ($h\nu$) and absorbance as a function of wavelength (inset) of a 500 nm thick CsI film deposited on a quartz substrate.}
  \label{l2ea4-f3}
\end{figure}

The UV/Vis absorption of a CsI film, deposited on a quartz substrate was measured in the spectral range 190 nm to 900 nm, as shown in the inset of Figure 2. An absorption peak is observed in the UV wavelength region (wavelength smaller than 225 nm) and its amplitude is found to be $\sim$3.5. The absorbance lies in the UV-region at a wavelength smaller than 225 nm. A similar result for absorbance spectra has been previously reported by Lu and McDonald \cite {C.Lu} for a 200 nm thick CsI film. The band gap of the photocathode is one of the key parameter determining the range of its most efficient operation, in particular the sensitivity cutoff. In addition to an appropriate optical band gap energy, a good photocathode material should allow an efficient electron transport to the emission surface and should have low or negative work function or electron affinity.

The absorption  of incident photons in the UV region is attributed to the band gap absorption of the CsI film. The large increase in the absorption for wavelengths smaller than 225 nm can be assigned to the intrinsic band gap absorption due to the electron excitation from the valence band to the conduction band. The absorption  band gap energy ($E_{g}$) has been calculated by using the  Tauc relation \cite{tauc1}.

\begin{equation}
  (\alpha h\nu)^{n}=A(h\nu-E_{g}),
\end{equation}

where $A$ is the edge width parameter, $h$ is the Planck's constant, $\nu$ is the frequency of vibration, $h\nu$ is the photon energy, $\alpha$ is the absorption coefficient, $E_{g}$ is the band gap energy and $n$ is either 2 for direct band transitions or 1/2 for indirect band transitions. The absorption coefficient $\alpha$ has been determined using the relation
~\cite{Poelman}:
\begin{equation}
   \alpha = \frac{1}{t}{\textrm{ln}}\frac{1}{T},
\end{equation}
where $t$ is the film thickness and $T$ is the optical transmittance. The optical band gap energy estimated from a Tauc plot of $(\alpha h \nu)^{2}$ versus photon energy $h\nu$ according to the K. M. Model~\cite{KM_model1,KM_model2} is shown in Figure 2. It has been observed that the plot of $(\alpha h \nu)^{2}$ vs. $h\nu$ is linear over a wide range of photon energies indicating a direct type of transitions. The value of photon energy $(h\nu)$ extrapolated to $\alpha$ = 0 gives an absorption edge which corresponds to the optical band gap energy $E_{g}$. The extrapolation gives an optical band gap energy $E_{g} \approx 5.43~eV$, which can be compared with the  band gap energy $E_{g} \approx 5.9~eV$ derived from experimental QE dependence on wavelength~\cite{buzulutskov} for heat enhanced CsI thick films.

\subsection*{Optical transmittance and refractive index}
\begin{figure}[hbpt]
  \centering
  \includegraphics*[width=65mm]{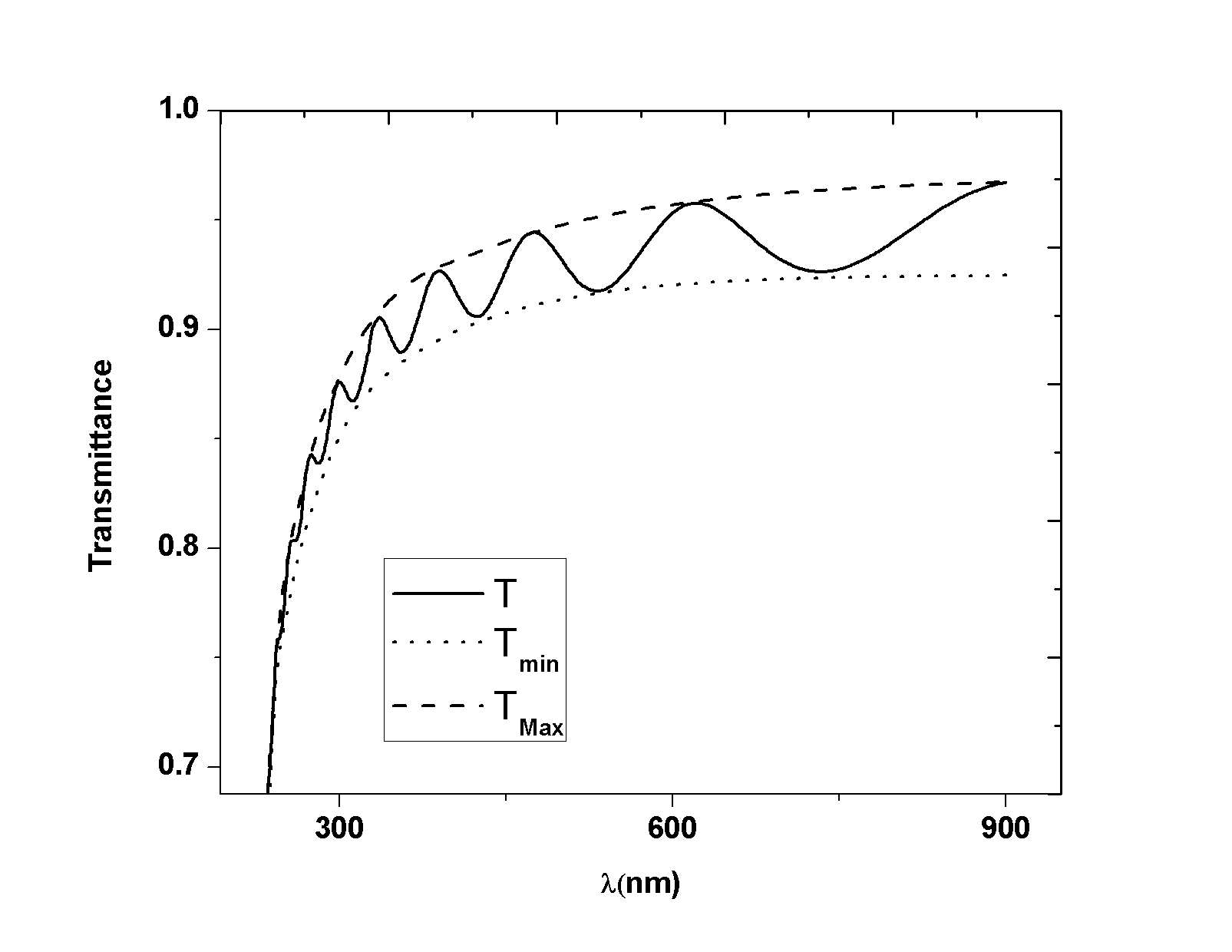}
  \caption{Transmission spectrum of a 500 nm thick CsI film (solid line), including the maximum $T_{max}$ and minimum $T_{min}$ transmittance envelope curves (dashed and dotted lines).}
  \label{l2ea4-f3}
\end{figure}

The optical transmittance of a 500 nm thick CsI photocathode  deposited on a quartz substrate has been measured in the wavelength range 190 nm to 900 nm. The transmittance is shown in Figure 3 (solid line). It has been calculated from the absorbance data using the relation:
\begin{equation}
  T=exp(-A),
\end{equation}
where A is the absorbance of the film. In the wavelength range 190 nm to 225 nm, the maximum transmittance is found to be be about 3.5\%, while in the wavelength range 225 nm to 900 nm, more than 80\% transmittance is observed. Consequently the CsI photocathode is opaque below 225 nm and transparent above 225 nm.

The surface quality and homogeneity of the CsI film was analyzed from the existence of interference fringes in the transmission spectra as shown in Figure 3. In the transparent spectral region ($\lambda \ge$ 225 nm), the CsI film shows an interference fringe pattern, which reveals the existence of continuous and homogeneous CsI layers. The spectrum shown in Figure 3 depicts a sharp fall in transmission near the fundamental absorption, which is an evidence for the good crystallinity of the film~\cite{sahay}. The oscillatory nature of the transmission spectrum is attributed to the interference of the light transmitted through the thin film and the substrate.

The optical properties of the CsI film can be evaluated from transmittance data using the method proposed by Swanepoel ~\cite{Swanepoel,manifacier}. The applicability of this method is limited to thin CsI films deposited on much thicker substrates. The application of this method entails, as a first step, the calculation of the maximum transmittance ($T_{max}$) and minimum transmittance ($T_{min}$) envelope curves by parabolic interpolation to the experimentally determined positions of peaks and valleys. From $T_{max}$ and $T_{min}$, the refractive index ($n_{\lambda}$) can then be calculated by using the expression ~\cite{manifacier}:

\begin{figure}[!ht]
  \begin{center}
    \includegraphics[scale=0.32]{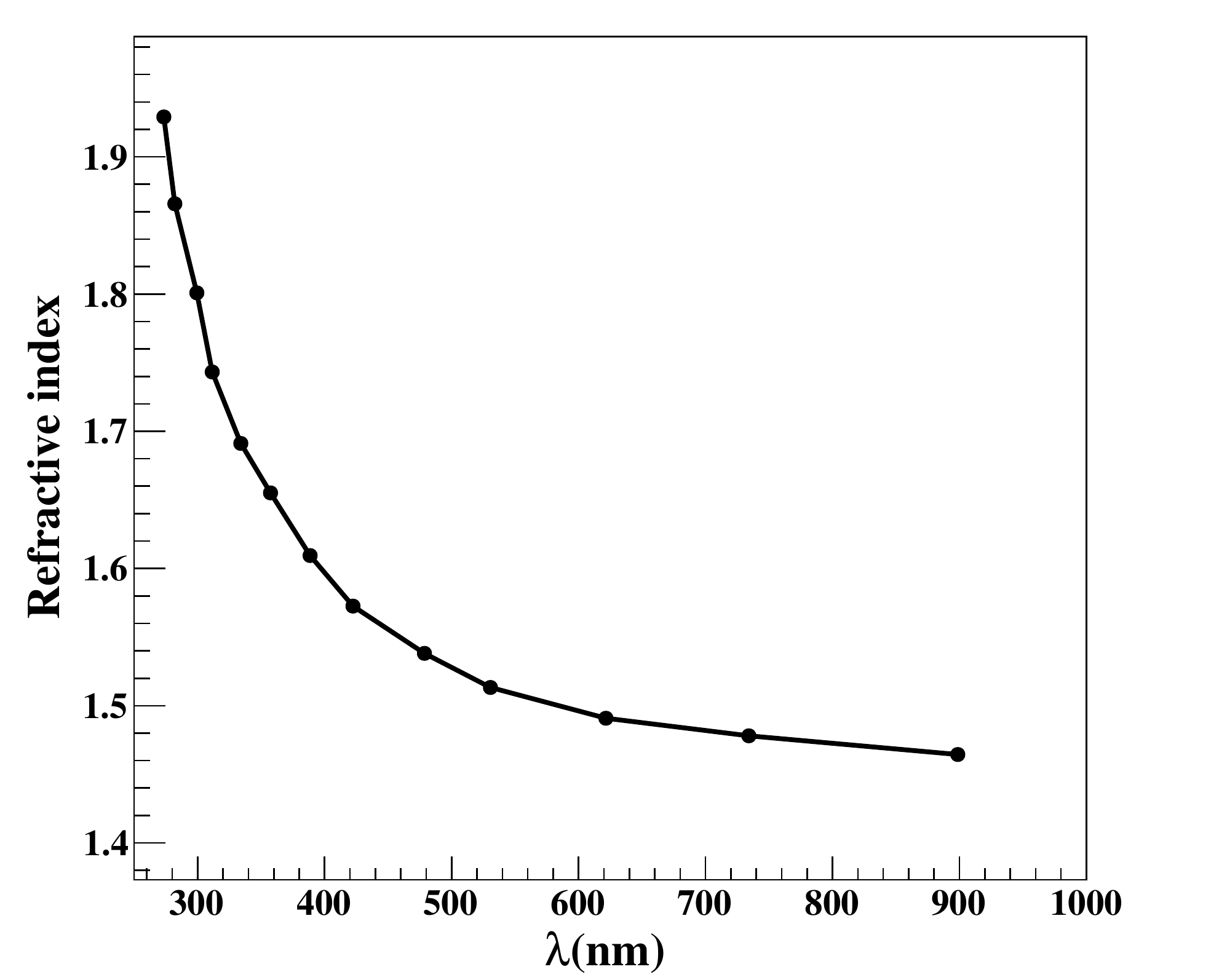}
    \caption{Refractive index as a function of wavelength for a 500 nm thick CsI film photocathode deposited on quartz  substrate. .}
    \label{fig1}
  \end{center}
\end{figure}

\begin{equation}
  n=\sqrt{\left [N+\sqrt{N^{2}-n_{s}^{2}}  \right ]}.
\end{equation}

In the weak and medium absorption regions, the value of N is given by:

\begin{equation}
  N=2n_{s}\frac{T_{max}-T_{min}}{T_{max}T_{min}}+\frac{n_{s}^{2}+1}{2},
\end{equation}

with $n_{s}$ being the refractive index of the substrate. In general, $n_{s}$ is determined by the maximum of the transmission in the transparent region $T_{max}$ ~\cite{minkov} using the relation:

\begin{equation}
  n_{s}= \frac{1}{T_{max}} + \sqrt{\left ( \frac{1}{T_{max}^{2}}-1 \right )}.
\end{equation}

The Swanepoel's envelope method~\cite{Swanepoel} is employed to estimate the refractive index from the transmittance spectra of 500 nm thick CsI film. It is observed that the value of the refractive index decreases with increasing wavelength (Figure 4), which shows that the film has a normal dispersive behavior.

\begin{figure}[hbpt]
  \centering
  \includegraphics*[width=60mm]{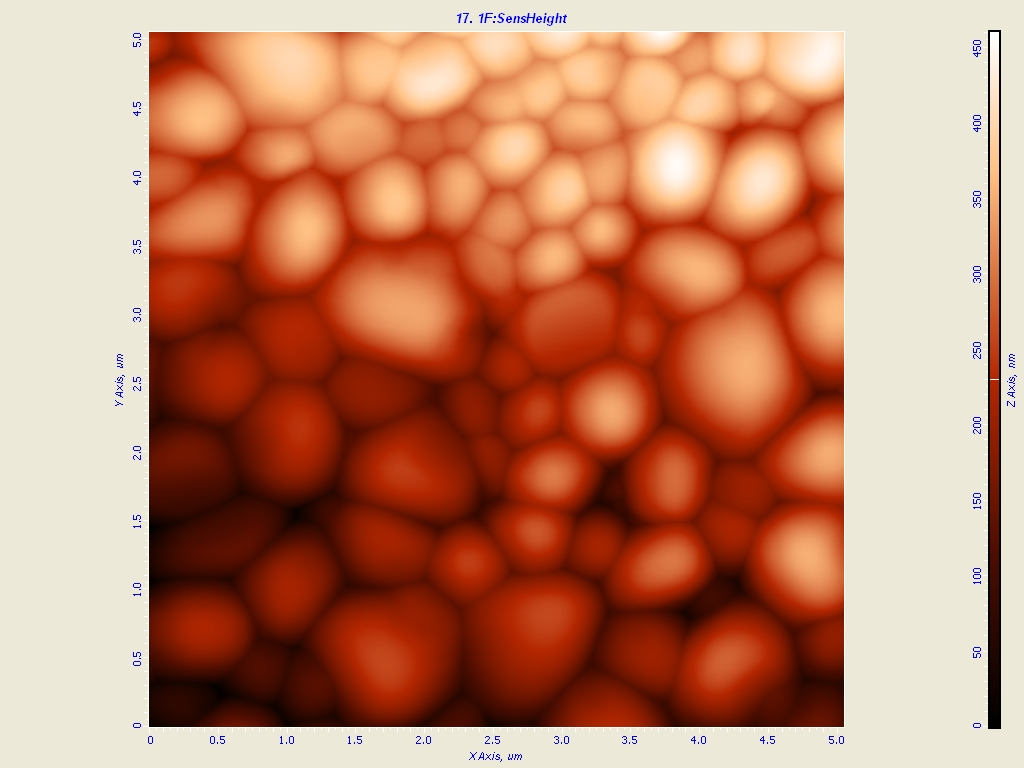}
  \includegraphics*[width=60mm]{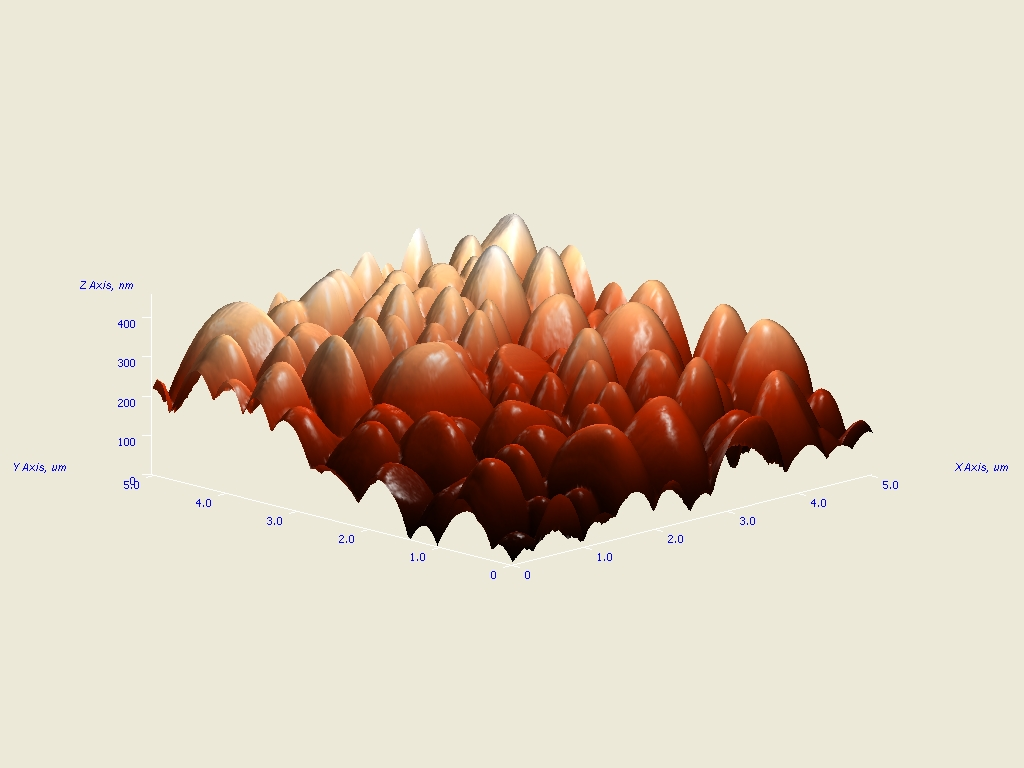}
  \caption{2-D (top panel) and 2-D (bottom panel) atomic force microscopy image of "as-deposited"  500 nm thick CsI film.}
  \label{l2ea4-f3}
\end{figure}

\section{Morphological properties of CsI}

In order to analyze the microscopic details of the surface structure of thin CsI films, such as grain size, average roughness and the arrangement of grains, AFM measurement has been adopted. AFM is a high-resolution type of scanning probe microscopy for morphological  information analysis. The AFM measurement technique provides digital images which allow quantitative measurements of surface features, such as root mean square (RMS) or average roughness and the analysis of images from different perspectives, including 2-D views. The top panel and bottom panel  respectively, of  Figure 5 show 2-D and (3-D) AFM surface images of 500 nm thick CsI film. The AFM surface images were recorded by scanning over an area of 5$\times$5 $\mu$$m^{2}$. The 2-D AFM surface image exhibits uniform film surface with pin hole free, crack free and densely packed morphology. The observed image shows clearly the grainy structure with ridge and valley features having an average roughness of about 39 nm and the grain size varying from 220 nm to 1700 nm. The low roughness of the film indicated that the film quality is good. The average grain size is calculated to be about 350 nm and the z-height is found to be about 170 nm. The morphological result obtained from AFM can be compared with earlier published results~\cite{triloki,triloki1}. The estimated average grain size (i.e. $\sim350~nm$) is comparable with earlier published result determined from transmission electron microscopy (i.e. $\sim300~nm$). 

\section{Elemental composition of CsI}

\begin{figure}[!ht]
  \begin{center}
    \includegraphics[scale=0.26]{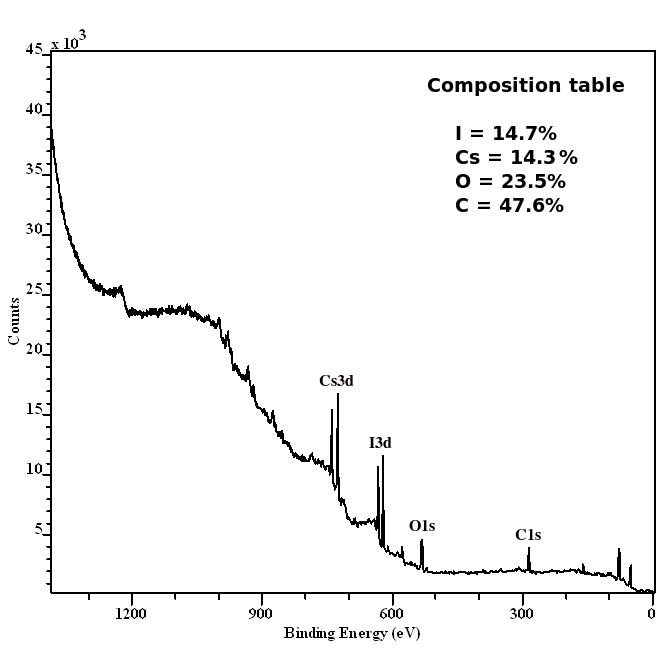}
    \caption{Elemental composition of the CsI film was determined by XPS. The spectrum suggests that the the film is mainly composed of Cs and I elements.}
    \label{comparision_thickness.eps}
  \end{center}
\end{figure}

The XPS technique was used to detect elements present in considerable amount (quantitative determination of bulk element composition). The XPS analysis was carried out on a 500 nm thick CsI film deposited on an Al substrate. Figure 6 shows the resulting XPS survey scan. The Cs3d and I3d transitions at a binding energy of 723.9 eV and 618.3 eV respectively (5/2 peaks of 3d spin-orbit doublet) are the dominant peaks. Carbon and Oxygen were also detected as a strong C1s signal at 285 eV and a weak O1s at 532 eV.  The signals of Carbon and Oxygen might be due to the atmospheric exposure during the sample transfer in the XPS chamber. The relative elemental concentrations are calculated from the peak intensities. The atomic ratio of I and Cs is found to be $\sim1.02$, indicating that the TE deposition results in a CsI film with almost stoichiometric composition. This is in good agreement with previous XPS results~\cite{bksingh3} using the same evaporation technique. However for CsI film deposited by pulse laser deposition (PLD) technique, ex situ XPS measurement shows an atomic ratio of I and Cs to be of relatively smaller value (0.52) as reported by Fairchild ~\cite{fairchild}, indicating a Cs-enriched surface.

\section{Conclusions}
The photoemission of thermally evaporated, 500 nm thick, CsI film was measured in the spectral range 150 nm to 200 nm. The maximum value of absolute QE is $\sim40\%$ at the wavelength $\lambda$ = 150 nm. The combined data from the optical measurements shows an optical band gap energy of about 5.43 eV in the UV spectral region with QE of about 6\% to 20\%.  The UV/Vis Optical data reveals that, CsI film is opaque in the spectral range of  190 nm to 225 nm, where as in the  spectral range of 225 nm to 900 nm, it is found to be almost transparent. The oscillatory nature in transmittance data shows the  existence of continuous and homogeneous CsI layers. The value of refractive index calculated from the envelope plot of optical transmittance data, varies from 1.82 to 1.30 in the spectral range 275 nm to 900 nm. The variation of refractive index indicates a dispersive behavior of the CsI film.  The AFM results reveal that the CsI films have crystalline, homogeneous and continuous grain like morphology. The average value of grain size and average film roughness was found to be about 350 nm and 39 nm respectively. XPS wide spectrum suggests that CsI film is mainly composed of Cs and I elements. The atomic ratio of Cs and I is found to be $\sim$1:1, which is consistent with the stoichiometry of CsI.

\section{Acknowledgment}
This work was partially supported by the Department of Science and Technology (DST), the Council of Scientific and Industrial Research (CSIR) and by the Indian Space Research Organization (ISRO), Govt. of India.  The authors would like to thank the referee who has contributed to make this paper more articulate.  Triloki and R. Rai acknowledge the University Grant Commission (UGC), New Delhi, India for providing financial support.


\begin{thebibliography}{00}
  
\bibitem{Tremsin}
  A. S. Tremsin, O. H. W. Siegmund, Nucl. Intr. and Meth. A 504 (2003) 4-8 and references therein.
  
\bibitem{breskin}
  A. Breskin, Nucl. Intr. and Meth. A 371 (1996) 116-136 and references therein.
  
\bibitem{bksingh1}
  B. K. Singh, E. Shefer, A. Breskinet, Nucl. Intr. and Meth. A 454 (2000) 364-378 and references therein.
	
\bibitem{bksingh}
  B. K. Singh, M. A. NItti, A. Valentini, Nucl. Intr. and Meth. A 581 (2007) 651-655 and references therein.
  
\bibitem{C.Lu}
  C. Lu and K. T. Mcdonald, Nucl. Intr. and Meth. A 343 (1994) 135-151.	
  
\bibitem{tauc1}
  J. Tauc, R. Grigorovici, and A. Vancu, Phys. Status Solidi, 15 (1966) 627.
  
\bibitem{Poelman}
  D. Poelman and P. F. Smet, J. Phys. D: Appl. Phys. 36 (2003) 1850-1857.
  
\bibitem{KM_model1}
  Yeong II Kim, Stephen J. Atherton, Elaine S. Brigham et al., J. Phys. Chem. 97 (1993) 11802-11810.

\bibitem{KM_model2}
Vesna D$\breve{z}$imbeg-Mal$\breve{c}$i$\acute{c}$, $\breve{Z}$eljka Barbari$\acute{c}$-Miko$\breve{c}$evi$\acute{c}$ and Katarina Itri$\acute{c}$, Technical Gazette 18, 1(2011) 117-124.

\bibitem{buzulutskov}
  A. Buzulutskov, A. Breskin and R. Chechik, J. Appl. Phys. 77 (1995) 2138-2145.
  
\bibitem{sahay}
  P. P. Sahay, R. K. Nath and S. Tewari, Cryst. Res. Technol. 42 (3) (2007) 275.
  
\bibitem{Swanepoel}
  R. J. Swanepoel, J. Phys. E: Sci. Instrum. 16 (1980) 1214.
  
\bibitem{manifacier}
  J. C. Manifacier, J. Gasiot and J P Fillard, J. Phys. E: Sci. Instrum. 9,(1976) 1002.
  
\bibitem{minkov}
  D. A. Minkov, J. Phys. D. Appl. Phys. 22 (1989) 199-205.
  
\bibitem{triloki}
  Triloki, B. Dutta and B. K. Singh, Nucl. Intr. and Meth. A 695 (2012) 279-282.
  
\bibitem{triloki1}
  Triloki, P. Garg, R. Rai, B. K. Singh et al., Nucl. Intr. and Meth. A 736 (2014) 128-134.
  
\bibitem{bksingh3}
  B. K. Singh, Triloki, P. Garg et al., Nucl. Intr. and Meth. A 610 (2009) 350-353.
  
\bibitem{fairchild}
  S. B. Fairchild, T. C. Back, P. T. Murray et al., J. of Vac. Sci. Technol. A 29, (2011) 031402(1-6).	
  
\end{thebibliography}


\end{document}